\documentclass[conference]{IEEEtran}
\IEEEoverridecommandlockouts
\usepackage{cite}
\usepackage{amsmath,amssymb,amsfonts}
\usepackage{algorithmic}
\usepackage{graphicx}
\usepackage{textcomp}
\usepackage{xcolor}
\usepackage{bm}
\usepackage{balance}

\newcommand{\vect}[1]{\mathbf{#1}}
\newcommand{\matr}[1]{\mathbf{#1}}
\newcommand{\herm}{\mathsf{H}}
\newcommand{\trans}{\mathsf{T}}
\newcommand{\E}{\mathbb{E}}
\newcommand{\diag}{\text{diag}}

\def\BibTeX{{\rm B\kern-.05em{\sc i\kern-.025em b}\kern-.08em
    T\kern-.1667em\lower.7ex\hbox{E}\kern-.125emX}}
\begin{document}

\title{Scale-Free Beamforming using Swarm Arrays for Remote Sensing under Interference \\
\thanks{This work is partially supported by NSF grant 2540120.}
}

\author{\IEEEauthorblockN{1\textsuperscript{st} Bradley Hamilton}
\IEEEauthorblockA{\textit{Electrical \& Computer Engineering} \\
\textit{University of Iowa}\\
Iowa City, USA \\
email address or ORCID}
\and
\IEEEauthorblockN{2\textsuperscript{nd} Raghu Mudumbai}
\IEEEauthorblockA{\textit{Electrical \& Computer Engineering} \\
\textit{University of Iowa}\\
Iowa City, USA \\
email address or ORCID}
\and
\IEEEauthorblockN{3\textsuperscript{rd} Soura Dasgupta}
\IEEEauthorblockA{\textit{Electrical \& Computer Engineering} \\
\textit{University of Iowa}\\
Iowa City, USA \\
email address or ORCID}
\and
\IEEEauthorblockN{4\textsuperscript{th} Benjamin Peiffer}
\IEEEauthorblockA{\textit{Applied Research and Technology} \\
\textit{Collins Aerospace}\\
Cedar Rapids, USA \\
benjamin.peiffer@collins.com}
}

\maketitle

\begin{abstract}
We consider a {\it swarm array} of autonomous relays that seek to cooperatively forward a desired signal to a fusion center with the maximum possible fidelity while canceling out a number of interferers. We present a distributed algorithm for computing the optimal zero-forcing beamforming weights at the relays without requiring prior channel knowledge. Crucially, our algorithm is {\it scale-free} in the sense that the computational and bandwidth overheads are completely independent of the size of the array. We build on recent work that introduced the concept of a Collective Array that enables such {\it scale-free} computation by imposing a constraint that the array must always function as a {\it swarm} i.e. array elements can only ever communicate with external nodes collectively and never individually. While this is a very severe restriction, we show that it allows useful computations such as zero-forcing beamforming while being robust to noise and channel time-variations.
\end{abstract}

\begin{IEEEkeywords}
scale-free arrays, cooperative relays, distributed beamforming, subspace projection, least mean squares
\end{IEEEkeywords}

\section{Introduction}
We consider the problem of optimally isolating a desired signal from multiple co-channel interfering signals using a distributed sensor array. We propose a novel {\it scale-free} method for solving this problem under which the computational and bandwidth overheads of optimal beamforming and interference cancellation are {\it entirely independent of the size of the sensor array}. This potentially allows massive scaling of such arrays and their deployment in situations where it is necessary to operate autonomously with limited coordination.

\begin{figure*}[htbp]
\centerline{\includegraphics[width=6in]{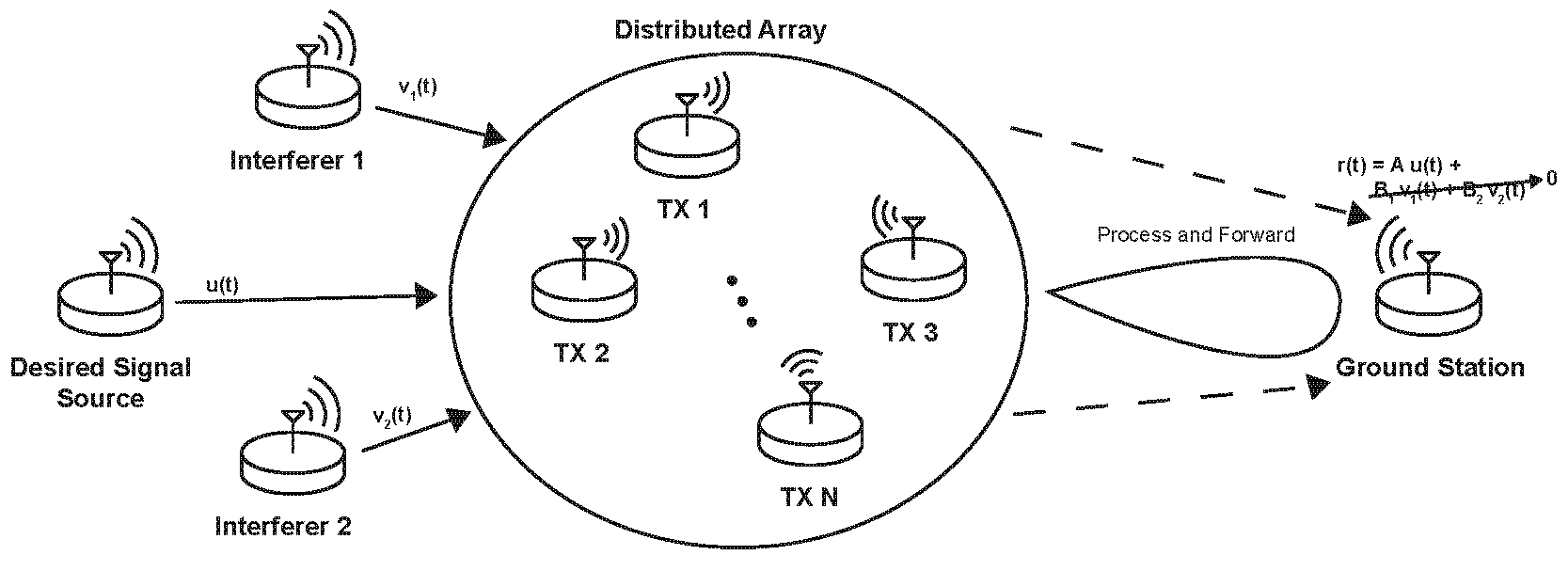}}
\caption{The swarm array concept enables a group of disadvantaged nodes to collectively mitigate interference and steer a receive beam toward a signal of interest by fusing their received signals at a central location. The swarm array always acts as a collective, ensuring no increase in coordination overhead as the array size scales to thousands of nodes.}
\label{fig:concept}
\end{figure*}

In our conception, the sensor array is deployed as a network of amplify-and-retransmit relays to construct a collaborative beamformer as shown in Fig. \ref{fig:concept}. Calculating the optimal amplification weights requires accounting for a cascaded, two-hop channel model: the sensing channel from the sources to the relay array, and the reachback channel from the relay array to the final destination (fusion center). The goal of sensor fusion can then be formalized as the problem of computing a complex weight vector that maximizes the received SNR of the desired source at the fusion center while enforcing spatial nulls (zero-forcing) against all known interferers. Our main contribution is a distributed, scale-free algorithm for adaptively computing these optimal weights in the sensor array.

\subsection{The Scale-Free Swarm Array}

We leverage the Collective Array concept introduced in \cite{collectivearray}. The Collective Array is defined by the constraint that all array processing must be {\it completely scale-free} in the sense that the computational and bandwidth requirements are entirely independent of the array size. This constraint is guaranteed by requiring that all outward communication from the array must only be in the form of aggregated signals i.e. combinations of signals from all individual sensors, while incoming communication is always addressed to the array as a whole and never to any individual element. All external entities only send and receive messages to the array as a collective, which in effect, renders individual elements to be entirely invisible to all external nodes and even to each other. 

This Collective Array concept ensures {\it scale-free} operation, but represents an extremely strong constraint. Crucially, the previous work in \cite{collectivearray} shows that it is possible to realize powerful capabilities under these extreme constraints. In concrete terms, \cite{collectivearray} provides a distributed algorithm for beamforming in a Collective Array that iteratively calculates a sequence of weight matrices converging to the optimal zero-forcing matrix without requiring any prior Channel State Information (CSI). This turns out to be a powerful building block for our sensor fusion problem.

\subsection{Contributions.}

\begin{itemize}
    \item \textbf{Equivalence to Scale-free Beamforming:} We show that under the assumption of dominant noise at the fusion center, the sensor fusion problem reduces to a receive beamforming task over an effective cascaded channel. We show that the subspace projection algorithm of \cite{collectivearray}, originally developed for centralized massive MIMO arrays, can be leveraged to achieve a distributed, scale-free solution for the relay-based sensor fusion problem.
    
    \item \textbf{Limitations of Subspace Projection:} We identify a critical limitation in geometric projection methods: thermal noise and finite precision results in irreversible subspace drift and the degradation of spatial nulls. We propose a new scale-free Normalized Least Mean Squared (NLMS) algorithm that mitigates these limitations of the subspace projection method.
    
    \item \textbf{Hybrid Acquisition/Tracking Operation:} We motivate a two-phase hybrid algorithm: utilizing subspace projections for rapid cold-start acquisition, followed by a smooth transition to NLMS-style tracking mode for robust steady-state maintenance.
\end{itemize}

\subsection{Background and Related Work.}

The basic ideas behind geometric projections and stochastic gradient descent for adaptive filtering goes back at least to the 1970s. In an influential contribution, Frost in 1972 \cite{frost1972} introduced the Linearly Constrained Minimum Variance (LCMV) beamformer that used spatial nulls as constraints. Originally designed for radar and sonar systems, Frost's algorithm utilized a geometric projection to dynamically null unknown interferers without distorting the desired ``look direction''. 

However, the limitations of linear constrained beamformers in the presence of numerical precision and noise were quickly recognized \cite{monzingo2004introduction} and researchers sought ways to relax the rigid geometric constraints \cite{griffiths1982alternative, zahm2007application}. A pivotal development was the Affine Projection Algorithm (APA) \cite{ozeki1984adaptive}, motivated by the slow convergence of standard LMS in the presence of highly correlated signals, such as human speech in acoustic echo cancellation. By mathematically projecting the weight updates onto a lower-dimensional subspace defined by recent data samples, APA established a formal theoretical connection between exact geometric solvers and Normalized LMS (NLMS). Combined with concurrent developments in regularized or ``leaky'' adaptive filters, these methods established that geometric projection methods must be ``softened'' to achieve steady-state stability.

While these earlier methods were designed for centralized processors, more recent work has adapted these principles for decentralized wireless relay networks \cite{bertrand2013distributed, lopes2008diffusion}. The stochastic channel estimation method described in \cite{goguri2016class} is well-suited for the scale-free array processing that we consider. However this method relies on statistical averaging and is slow compared to the subspace projection method introduced in \cite{collectivearray}.

{\bf Outline.} The rest of the paper is organized as follows. A formal mathematical statement of the sensor fusion problem is presented in Section \ref{sec:prob}. A distributed solution suitable for Collective Arrays is presented in Section \ref{sec:dist} and its performance in the presence of noise and channel time-variations is presented in Section \ref{sec:dynamic}. Section \ref{sec:results} illustrates the theoretical claims with numerical simulations and Section \ref{sec:conc} concludes.

\section{Problem Statement} \label{sec:prob}

Consider an environment with $M$ single-antenna transmitters and a receiving array of $N$ elements. Without loss of generality, let transmitter $1$ be the desired source and transmitters $2$ through $M$ be the interfering sources. Let the transmitted signals be defined by the $M \times 1$ vector $\vect{s}(t)$:
\begin{equation}
    \vect{s}(t) = [s_1(t), s_2(t), \dots, s_M(t)]^\trans
\end{equation}

The signals propagate through a narrowband, flat-fading channel to the $N$-element array. The baseband signal received at the array is given by:
\begin{equation}
    \vect{r}(t) = \matr{H}\vect{s}(t) + \vect{n}_r(t)
\end{equation}
where $\matr{H}$ is the $N \times M$ complex channel matrix, and $\vect{n}_r(t)$ is the $N \times 1$ additive noise vector at the relay array. We assume the relay noise is spatially white with covariance $\E[\vect{n}_r(t)\vect{n}_r^\herm(t)] = \sigma_r^2 \matr{I}$.

\subsection{Amplify-and-Retransmit Architecture}
Each of the $N$ relay elements applies a complex scalar weight to its received signal before retransmitting it to the fusion center. To facilitate standard beamforming notation, we assume the array elements apply the complex conjugate of the weight vector $\vect{w} \in \mathbb{C}^{N \times 1}$. The transmitted signal vector from the array is:
\begin{equation}
    \vect{x}(t) = \diag(\vect{w}^*)\vect{r}(t)
\end{equation}

The retransmitted signals propagate to the fusion center through an $N \times 1$ channel vector $\vect{g} = [g_1, g_2, \dots, g_N]^\trans$. The composite signal received at the fusion center is:
\begin{equation}
    c(t) = \vect{g}^\trans \vect{x}(t) + n_c(t) = \vect{g}^\trans \diag(\vect{w}^*) \vect{r}(t) + n_c(t)
\end{equation}
where $n_c(t)$ is the noise at the fusion center. Recognizing that $\vect{g}^\trans \diag(\vect{w}^*) = \vect{w}^\herm \diag(\vect{g})$, we rewrite the cascaded system as:
\begin{equation}
    c(t) = \vect{w}^\herm \diag(\vect{g}) \matr{H}\vect{s}(t) + \vect{w}^\herm \diag(\vect{g})\vect{n}_r(t) + n_c(t)
\end{equation}

\subsection{Zero-Forcing Beamforming Formulation}
We define an effective cascaded channel matrix $\matr{V} \in \mathbb{C}^{N \times M}$ that combines the sensing and reachback channels:
\begin{equation}
    \matr{V} = \diag(\vect{g})\matr{H}
\end{equation}
The term $\vect{w}^\herm \matr{V} \vect{s}(t)$ represents the combined signal at the fusion center with the weights $\vect{w}$. We impose two simultaneous constraints on the weight vector:
\begin{enumerate}
    \item Unity gain for the desired signal: $\vect{w}^\herm \vect{v}_1 = 1$
    \item Zero-forcing for the $M-1$ interferers: $\vect{w}^\herm \vect{v}_m = 0 \quad \forall m \in \{2, \dots, M\}$
\end{enumerate}
These conditions can be compactly written using an $M \times 1$ constraint vector $\vect{e} = [1, 0, \dots, 0]^\trans$:
\begin{equation}
    \matr{V}^\herm \vect{w} = \vect{e}
\end{equation}
By fixing the gain of the desired signal and perfectly nulling the interferers, maximizing the end-to-end SNR is mathematically equivalent to minimizing the total noise power at the fusion center.

\subsection{Optimization Regimes and System Stability}

The optimal beamforming solution depends on the where the dominant noise source is in the system. 

\subsubsection{The Relay-Noise Limited Regime}
A significant analytical simplification arises if we assume that the channel to the fusion center operates at a high SNR, rendering $n_c(t)$ negligible. The total system noise is thus dominated by the amplified array noise:
\begin{equation} \label{eq:amplified_noise}
    P_N = \sigma_r^2 \sum_{n=1}^N |w_n|^2 |g_n|^2
\end{equation}
To minimize $P_N$, we can introduce an effective weight vector $\tilde{\vect{w}} = \diag(\vect{g}^*)\vect{w}$, which simplifies the objective to $\min \|\tilde{\vect{w}}\|^2$ subject to the isolated source-channel constraint $\matr{H}^\herm \tilde{\vect{w}} = \vect{e}$. The physical weights applied by the array are then obtained by local channel inversion: $w_n = \tilde{w}_n / g_n^*$. This channel inversion, however, introduces significant vulnerabilities such as:
\begin{enumerate}
    \item \textbf{Numerical Instability:} Because the objective function lacks a penalty on the total transmit power, a relay element entering a deep fade ($g_n \to 0$) is commanded to supply an asymptotically infinite physical weight causing a massive amplification of local thermal noise.
    \item \textbf{Null Degradation:} The zero-forcing algorithm operates obliviously to the reachback channel. Any error in a node's local reachback estimate $\hat{g}_n$ alters the transmitted signal phase. Because deep spatial nulls rely on perfect destructive interference, these local phase errors act as unsuppressed interference multipliers, thereby degrading the SINR.
\end{enumerate}

\subsubsection{The Fusion-Noise Limited Regime (Effective Channel Zero-Forcing)}
To guarantee system stability, we reformulate the objective under the assumption of a fusion-noise limited regime ($\sigma_c^2 \gg P_N$). In this regime, the noise floor is decoupled from the array's amplification weights, and maximizing system SNR reduces to maximizing power efficiency: satisfying the exact cascaded constraints while minimizing the aggregate physical transmit power of the swarm. 

This is formulated as minimizing the norm of the physical weights directly against the cascaded channel:
\begin{equation}
    \min_{\vect{w}} \|\vect{w}\|^2 \quad \text{subject to} \quad \matr{V}^\herm \vect{w} = \vect{e}
\end{equation}
The optimal weight vector is given by the Moore-Penrose right pseudoinverse of the cascaded channel matrix $\matr{V}^\herm$:
\begin{equation} \label{eq:zf_cascaded_opt}
    \vect{w}_{\text{opt}} = \matr{V} (\matr{V}^\herm \matr{V})^{-1} \vect{e}
\end{equation}

By zero-forcing the effective cascaded channel $\matr{V}$ directly, the physical weight assigned to node $n$ naturally scales proportionally to its reachback channel gain ($w_n \propto g_n$). If a relay experiences a deep fade, its weight organically goes to zero.

While the Relay-Noise Limited Regime presents an interesting technical challenge, its potential instabilities present challenges, a detailed consideration of which we defer to future work. We will focus on the simpler Fusion-Noise Limited Regime for the rest of the paper.

\section{Distributed Solution via the Collective Array Framework} \label{sec:dist}

We will now show a procedure to compute the optimal zero-forcing beamforming weights in \eqref{eq:zf_cascaded_opt} for the Fusion-Noise Limited Regime using the Collective Array framework.

\subsection{Native Projections via Over-the-Air Superposition}

The key observation is that our cooperative Amplify-and-Retransmit (A\&R) relay problem reduces exactly to beamforming in the Collective Array as defined in \cite{collectivearray}. Crucially, because we are zero-forcing the cascaded channel $\matr{V}$, the physical over-the-air superposition of the electromagnetic waves naturally acts as the distributed fronthaul required by the algorithm, entirely bypassing the need for explicit local channel estimation.

At iteration $k$, each array element $n$ independently generates a random complex trial weight $u_n[k]$. The algorithm requires measurement of the projection $\vect{u}^\herm[k] \matr{V}$. The physical environment computes this quantity automatically. When the array elements apply their trial weights $u^*_n[k]$ to the received signals and retransmit them, the waves propagate through the physical reachback channel gains $g_n$. The fusion center receives the physical superposition:
\begin{equation}
    y[k] = \sum_{n=1}^N g_n u^*_n[k] r_n(t)
\end{equation}
By correlating this received aggregate signal against the $M$ known source pilots, the fusion center directly measures the cross-talk vector:
\begin{equation}
    \vect{y}[k] = \matr{V}^\herm \vect{u}[k]
\end{equation}

Upon receiving the broadcast of this aggregate measurement $\vect{y}[k]$, each relay element $n$ evaluates its localized orthogonal trial weight $u_n^\perp[k] = u_n[k] - \vect{w}_n[k-1] \vect{y}[k]$, where $\vect{w}_n[k-1]$ represents the element's current $1 \times M$ physical weight vector. Using two additional aggregate quantities broadcast by the array processor---the correlation vector $\vect{\rho}[k]$ and the normalization scalar $\|\vect{u}^\perp[k]\|^2$---the element updates its weights using the projection:
\begin{equation}
    \vect{w}_n[k] = \vect{w}_n[k-1] - \frac{u_n^\perp[k] \vect{\rho}^\herm[k]}{\|\vect{u}^\perp[k]\|^2}
\end{equation}

Thus, the distributed range-space projection algorithm can iteratively determine the optimal physical weights $\vect{w}_{\text{opt}}$ using only the aggregate measurements naturally provided by the physical channel. The relay elements do not need to estimate, track, or invert their local fusion channel gains $g_n$; they simply generate trial weights, transmit, and update their states based on the feedback broadcast by the fusion center.
\section{Subspace Maintenance and Dynamic Tracking} \label{sec:dynamic}

The distributed zero-forcing procedure derived above provably converges to the optimal solution under ideal conditions. However, in practice, maintaining these  spatial nulls over time requires continuous adaptation due to two distinct mechanisms of subspace degradation: internal algorithmic drift driven by thermal noise, and external physical drift driven by mobility.

\subsection{Noise-Driven Subspace Drift}

We first consider the inherent stochastic behavior of the algorithm in a perfectly static environment where the cascaded channel $\matr{V}$ is constant. At each iteration $k$, the aggregate signal measured at the fusion center is corrupted by additive receiver noise, $n_c[k]$. Because the projection algorithm strictly restricts the updated weight matrix to the orthogonal complement of this measured signal, thermal noise is mathematically embedded into the new null space geometry. 

Rather than averaging out over time, these microscopic, noisy projections permanently remove valid degrees of freedom, causing the weight matrix $\matr{W}[k]$ to undergo a random walk. Consequently, this noise-driven subspace drift guarantees that the interference nulls will eventually degrade, causing a constraint violation: $\matr{W}^\herm[k] \matr{V} \neq \matr{I}_M$.

\subsection{Unified Tracking of the Cascaded Relay Channel}

In practice, this algorithmic drift is compounded by the macroscopic mobility of the sources or the sensor swarm. If either the sensing channel $\matr{H}$ or the reachback channel $\vect{g}$ is time-varying, the zero-forcing target also changes.

In a naive relay beamforming approach, where a swarm must explicitly maintain estimates of both the source channels $\matr{H}$ and the reachback channels $\vect{g}$, the inversion of local gains (i.e., $w_n = \tilde{w}_n / \hat{g}_n^*$) means that any phase or amplitude error in the local reachback estimate $\hat{g}_n$ acts as a massive unsuppressed interference multiplier. This requires some regularization mechanism and consequent tradeoffs between performance and robustness.

Fortunately, because we are directly zero-forcing the cascaded channel matrix $\matr{V}(t) = \diag(\vect{g}(t))\matr{H}(t)$, the physical environment naturally combines the variation of both channels into a single effective channel. A phase drift due to a moving source (changing $\matr{H}$) or a moving relay (changing $\vect{g}$) manifests identically at the fusion center: as an aggregate physical leakage error $\vect{e}[k]$. We will now consider several candidate methods for tracking these aggregate errors.

\subsection{Baseline: Periodic Batch Renormalization}

The problem of subspace drift under noise was recognized in \cite{collectivearray} and a simple solution was proposed: periodically reimpose the zero-forcing constraint. Suppose the array measures the current channel projections $\mathbf{Y}[k-1] \doteq \mathbf{G}^H[k-1] \mathbf{H}$. By setting $\mathbf{G}[k] = \mathbf{G}[k-1] \mathbf{Y}^{-1}[k-1]$, the constraint $\mathbf{G}^H[k] \mathbf{H} = \mathbb{I}_M$ is restored while the range space of the weights remains unchanged. 

While this method preserves scale-free operation, we will show that it has significant drawbacks. First, the ``batch'' nature of the update requires periodic matrix inversions and synchronized feedback, which adds discrete computational bursts to the array processor. Secondly and more fundamentally, the renormalization is ``subspace rigid''; it merely restores the constraint within the existing column space, offering no mechanism to discover new channel directions as the physical geometry rotates. 

\subsection{Tracking using LMS-like updates} \label{sec:lms}

The classical Least Mean Squares (LMS) algorithm is known to lend itself to distributed implementations, which motivates the use of LMS-like updates as an alternative to the Periodic Batch Renormalization. Suppose that at step $k$, the fusion center generates and broadcasts a shared random combination vector $\mathbf{q}[k] \in \mathbb{C}^{M \times 1}$. Let $\mathbf{w}_{n}[k]$ denote the $1 \times M$ physical weight row vector stored locally at relay element $n$.

Crucially, to ensure the array can discover new channel directions and is not mathematically trapped in its current subspace, each element adds to $\mathbf{q}[k]$ a random exploration term $\delta_n[k]$ to generate a trial weight $u_n[k]$ for the current iteration:
\begin{equation}
    u_n[k] = \mathbf{w}_{n}[k] \mathbf{q}[k] + \alpha \delta_n[k]
\end{equation}
where $\alpha$ is a small exploration step-size parameter.

The relay array applies the complex conjugate of these trial weights, $u_n^*[k]$, to the received signals and transmits. The fusion center measures the effective $M \times 1$ leakage vector $\mathbf{y}[k]$ against the $M$ source pilots over the air. Because the signal propagates through the physical reachback channel, this measurement represents the true, instantaneous leakage of the cascaded system. The fusion center calculates the geometric error vector:
\begin{equation}
    \mathbf{e}[k] = \mathbf{y}[k] - \mathbf{q}[k]
\end{equation}

The fusion center broadcasts this aggregate error $\mathbf{e}[k]$. Each array element $n$ then performs a continuous gradient descent update on its local weight vector with step size $\mu$:
\begin{equation} \label{eq:lmsupdate}
    \mathbf{w}_{n}[k+1] = \mathbf{w}_{n}[k] - \mu (\mathbf{w}_{n}[k] \mathbf{e}[k]) \mathbf{q}^H[k]
\end{equation}
{\bf Remark.} Equation \eqref{eq:lmsupdate} is obtained by using $\mathbf{w}_n[k]$ as a stochastic proxy for the $n$-th row of $\mathbf{V}$ in the exact expression for the gradient of the MSE cost function (because of course the true channel gains $\mathbf{V}$ are unknown). Thus, this specific update rule will not work from a cold-start initialization: we need to first get the weights $\mathbf{w}_n[k]$ at least approximately aligned with the true channel gains to allow the LMS gradient search to work correctly.

The resulting update rule \eqref{eq:lmsupdate} obeys the scale-free constraints of the Collective Array: the correction term is computed entirely locally at each relay element $n$. Specifically, the inner product $(\mathbf{w}_{n}[k] \mathbf{e}[k])$ reduces to a scalar quantity requiring only $M$ local multiplications, where $M$ is the number of spatial dimensions. To perform this update, the fusion center only needs to broadcast the $M \times 1$ error vector $\mathbf{e}[k]$ and the $M \times 1$ pilot combination vector $\mathbf{q}[k]$. Because the fusion center measures the aggregate physical leakage $\mathbf{y}[k]$ directly over the air, it neither requests nor requires knowledge of the individual reachback channels or the local physical weights. Consequently, both the computational complexity at each node and the aggregate communication overhead are completely independent of the array size $N$.

Because the trial weights now contain the independent exploration term $\delta_n[k]$, the stochastic gradient naturally incorporates geometric components outside the array's current column space. Over time, the Law of Large Numbers averages out the independent random variables, allowing the gradient to smoothly rotate the physical weights to track variations in the cascaded channel $\mathbf{V}(t)$. This update requires only a single aggregate pilot measurement per iteration, entirely eliminates matrix inversions, and dynamically nudges the zero-forcing weights to counteract both thermal drift and physical mobility.

\section{Numerical Simulations} \label{sec:results}

We identify three combinations of the previous ideas that we will explore to understand the tradeoffs between fast convergence and robustness to noise and channel drifts.
\begin{itemize}
    \item \textbf{Method (1): Periodic Batch Renormalization.} The baseline approach where the constraint is periodically reset using the measured projection matrix $\mathbf{Y}$.
    
    \item \textbf{Method (2): Interleaved Subspace-LMS.} Alternating between subspace projection and NLMS updates to ``nudge'' the weight vector out of its subspace without requiring full batch resets.
    
    \item \textbf{Method (3): Hybrid Phased Approach.} Utilizing the high-speed range-space projection algorithm for initial acquisition and then switching to pure LMS for steady-state tracking of the converged subspace solution. In practice, this method should be augmented with logic that resets to the initial acquisition algorithm when sudden changes are discovered.
\end{itemize}

We hypothesize that while Method (1) will remain stable in static environments, its performance will be capped by an interference floor in mobile scenarios because it cannot track channel rotations outside its current span. Methods (2) and (3) represent two different attempts to balance between the subspace projection and NLMS approaches. We now test these hypotheses using numerical simulations.

\begin{figure}[ht!]
        \centering
        \includegraphics[width=\linewidth]{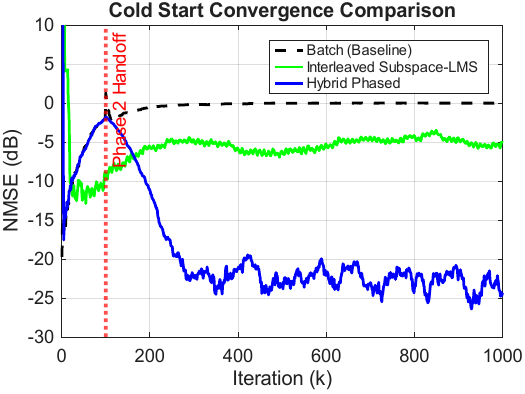}
        \caption{NMSE Vs Iterations with $K_{\text{batch}}=5,N=200,M=4, \text{SNR} =10 \text{ dB}$ and $T_{\text{switch}}=100$. By employing a two-phase algorithm with distinct acquisition and tracking modes, the hybrid phased approach achieves the same rapid convergence as other methods while improving steady-state performance by close to $25$ dB.}
        \label{fig:ColdStartFig}    \end{figure}

{\bf Convergence from Cold Start.} Figure \ref{fig:ColdStartFig} plots the normalized mean squared error (NMSE) error defined as $\frac{\|\mathbf{G}^H[k] \mathbf{H} -  \mathbb{I}_M \|^2}{\| \mathbf{G}^H[k] \mathbf{H} \|^2}$ as a function of iterations $k$ for a static channel of each of the three methods described above with a random initialization. In the hybrid method 3, the algorithm switches to the NLMS updates \eqref{eq:lmsupdate} after $100$ iterations. This provides very fast convergence as well as superior robustness to noise, and in low SNR ($\sim$ 10 dB), it achieves $\approx 20$ dB smaller steady-state NMSE compared to the other two  methods.

\begin{figure}[ht!]
        \centering
        \includegraphics[width=\linewidth]{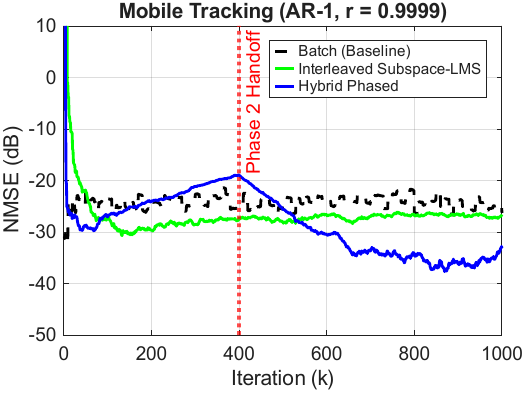}
        \caption{NMSE Vs Iterations with $K_{\text{batch}}=10,N=200,M=4, \text{SNR} =30 \text{ dB}$ and $T_{\text{switch}}=200$. Note that the hybrid phased approach achieves fast acquisition but starts drifting until the NLMS update handoff happens, at which point, it maintains superior tracking performance over the other two methods.}
        \label{fig:ARFig}
    \end{figure}

{\bf Tracking Slowly Time-Varying Channels.} In Figure \ref{fig:ARFig}, the channels vary in time according to a simple auto-regressive model with correlation parameter $r=0.9999$ per iteration. Note the relatively high SNR of $30$ dB in this simulation: we want to explore the effect of time-variation rather than noise. We again see the hybrid method 3 outperforming the other two.

\begin{figure}[hb!]
        \centering
        \includegraphics[width=\linewidth]{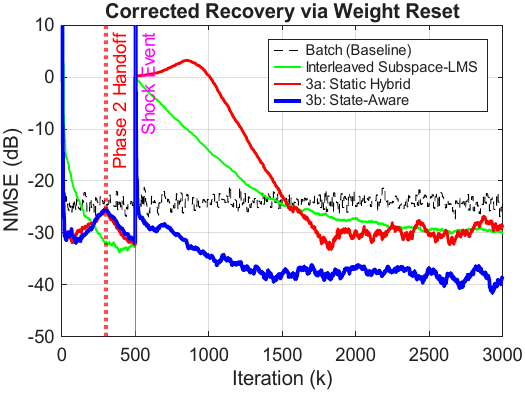}
        \caption{NMSE Vs Iterations with $K_{\text{batch}}=10,N=200,M=4, \text{SNR} =30 \text{ dB}$ and $T_{\text{switch}}=300$. By allowing the hybrid algorithm to re-initiate its acquisition process after a disturbance is detected, recovery time can be reduced to just a few iterations. The ability to rapidly recover is of utmost importance in scenarios where the dynamics of the swarm array may rapidly change.}
        \label{fig:SuddenChangeFig}
    \end{figure}

{\bf Adapting to Sudden Channel Change.} Figure \ref{fig:SuddenChangeFig} models how each method responds to a sudden change in the channel. In this simulation, we consider two variants of the hybrid method 3: Method 3a simply allows the LMS update to eventually ``catch up'' to the changed channel, whereas Method 3b has special logic to detect the sudden change and revert to the subspace projection. We see that this latter method combines the best of both the subspace and LMS updates: not only is it faster to adapt than the pure LMS updates in Method 3a, it also achieves lower steady-state NMSE.
    
\balance

\section{Conclusions} \label{sec:conc}

We introduced a distributed algorithm for optimal zero-forcing beamforming from a {\it swarm array} of relays to a fusion center without any prior channel knowledge. We leveraged the subspace projection algorithm presented in \cite{collectivearray} that has the property of being {\it scale-free} in the sense that its computational and bandwidth requirements are completely independent of array size. We showed that the subspace projection method performs well under ideal conditions, but lacks robustness to noise and channel drifts. We presented an NLMS-like algorithm that also satisfies the {\it scale-free} property and is robust to noise and time-variations, but needs some channel state information. We show by numerical simulations that a hybrid combination of the subspace projection and NMLS algorithms performs well under various conditions. A full theoretical understanding of these algorithms and limitations is a topic for future work. Exploring other capabilities that can be realized with scale-free arrays is another interesting open question.

\bibliographystyle{ieeetr}
\bibliography{refs}

@inproceedings{collectivearray,
  title="{Scalable Beamforming for Very Large Antenna Arrays without CSI}",
  author={Hamilton, Bradley and Mudumbai, Raghu and Dasgupta, Soura},
  booktitle={ICASSP 2026-2026 IEEE International Conference on Acoustics, Speech and Signal Processing (ICASSP)},
  pages={21466--21470},
  year={2026},
  organization={IEEE}
}

@inproceedings{goguri2016class,
  title="{A class of scalable feedback algorithms for beam and null-forming from distributed arrays}",
  author={Goguri, Sairam and Peiffer, Ben and Mudumbai, Raghu and Dasgupta, Soura},
  booktitle={2016 50th Asilomar Conference on Signals, Systems and Computers},
  pages={1447--1451},
  year={2016},
  organization={IEEE}
}

@ARTICLE{frost1972,
  author={Frost, O.L.},
  journal={Proceedings of the IEEE}, 
  title={An algorithm for linearly constrained adaptive array processing}, 
  year={1972},
  volume={60},
  number={8},
  pages={926-935},
  doi={10.1109/PROC.1972.8817}
  }

@book{monzingo2004introduction,
  title={Introduction to adaptive arrays},
  author={Monzingo, Robert A and Miller, Thomas W},
  year={2004},
  publisher={Scitech publishing}
}

@article{griffiths1982alternative,
  title={An alternative approach to linearly constrained adaptive beamforming},
  author={Griffiths, Lloyd and Jim, CW},
  journal={IEEE Transactions on antennas and propagation},
  volume={30},
  number={1},
  pages={27--34},
  year={1982},
  publisher={IEEE}
}

@article{zahm2007application,
  title={Application of adaptive arrays to suppress strong jammers in the presence of weak signals},
  author={Zahm, Charles L},
  journal={IEEE Transactions on Aerospace and Electronic Systems},
  number={2},
  pages={260--271},
  year={2007},
  publisher={IEEE}
}

@article{ozeki1984adaptive,
  title={An adaptive filtering algorithm using an orthogonal projection to an affine subspace and its properties},
  author={Ozeki, Kazuhiko and Umeda, Tetsuo},
  journal={Electronics and Communications in Japan (Part I: Communications)},
  volume={67},
  number={5},
  pages={19--27},
  year={1984},
  publisher={Wiley Online Library}
}

@article{bertrand2013distributed,
  title="{Distributed LCMV beamforming in a wireless sensor network with single-channel per-node signal transmission}",
  author={Bertrand, Alexander and Moonen, Marc},
  journal={IEEE Transactions on Signal Processing},
  volume={61},
  number={13},
  pages={3447--3459},
  year={2013},
  publisher={IEEE}
}

@article{lopes2008diffusion,
  title={Diffusion least-mean squares over adaptive networks: Formulation and performance analysis},
  author={Lopes, Cassio G and Sayed, Ali H},
  journal={IEEE Transactions on Signal Processing},
  volume={56},
  number={7},
  pages={3122--3136},
  year={2008},
  publisher={IEEE}
}
\end{document}